\documentclass[aps,prl,showpacs,twocolumn,superscriptaddress,longbibliography]{revtex4-2}

\usepackage{url}
\usepackage[colorlinks, citecolor=blue, linkcolor=blue, urlcolor=blue, breaklinks]{hyperref}
\usepackage[utf8]{inputenc}
\usepackage[pdftex]{graphicx}
\usepackage[table]{xcolor}

\usepackage[version=4]{mhchem}
\usepackage{physics}
\usepackage{amssymb}
\usepackage{array} 
\usepackage{changes}

\usepackage{multirow}
\usepackage{makecell}
\newcolumntype{s}{>{\columncolor[RGB]{220,220,220}} c}
\newcolumntype{P}[1]{>{\centering\arraybackslash}p{#1}}
\newcolumntype{M}[1]{>{\centering\arraybackslash}m{#1}}

\begin{document}
\title{Electronic structure of few-layer black phosphorus from $\mu$-ARPES}

\author{Florian Margot}
\affiliation{Department of Quantum Matter Physics, University of Geneva, 24 quai Ernest Ansermet, CH-1211 Geneva, Switzerland}
\author{Simone Lisi}
\affiliation{Department of Quantum Matter Physics, University of Geneva, 24 quai Ernest Ansermet, CH-1211 Geneva, Switzerland}
\author{Irène Cucchi}
\affiliation{Department of Quantum Matter Physics, University of Geneva, 24 quai Ernest Ansermet, CH-1211 Geneva, Switzerland}
\author{Edoardo Cappelli}
\affiliation{Department of Quantum Matter Physics, University of Geneva, 24 quai Ernest Ansermet, CH-1211 Geneva, Switzerland}
\author{Andrew Hunter}
\affiliation{Department of Quantum Matter Physics, University of Geneva, 24 quai Ernest Ansermet, CH-1211 Geneva, Switzerland}
\author{Ignacio Gutiérrez-Lezama}
\affiliation{Department of Quantum Matter Physics, University of Geneva, 24 quai Ernest Ansermet, CH-1211 Geneva, Switzerland}
\affiliation{Group of Applied Physics, University of Geneva, 24 quai Ernest Ansermet, CH-1211 Geneva, Switzerland}
\author{KeYuan Ma}
\affiliation{Department of Chemistry, University of Zürich, CH-8057 Zürich, Switzerland}
\author{Fabian von Rohr}
\affiliation{Department of Quantum Matter Physics, University of Geneva, 24 quai Ernest Ansermet, CH-1211 Geneva, Switzerland}
\author{Christophe Berthod}
\affiliation{Department of Quantum Matter Physics, University of Geneva, 24 quai Ernest Ansermet, CH-1211 Geneva, Switzerland}
\author{Francesco Petocchi}
\affiliation{Department of Quantum Matter Physics, University of Geneva, 24 quai Ernest Ansermet, CH-1211 Geneva, Switzerland}
\author{Samuel Ponc\'e}
\affiliation{Institute of Condensed Matter and Nanosciences, Universit\'e catholique de Louvain, BE-1348 Louvain-la-Neuve, Belgium}
\author{Nicola Marzari}
\affiliation{Laboratory of theory and simulation of materials, École Polytechnique Fédérale de Lausanne, CH-1015 Lausanne, Switzerland}
\author{Marco Gibertini}
\affiliation{Dipartimento di Scienze Fisiche, University of Modena and Reggio Emilia, Modena, Emilia-Romagna, Italy}
\author{Anna Tamai}
\affiliation{Department of Quantum Matter Physics, University of Geneva, 24 quai Ernest Ansermet, CH-1211 Geneva, Switzerland}
\author{Alberto F. Morpurgo}
\affiliation{Department of Quantum Matter Physics, University of Geneva, 24 quai Ernest Ansermet, CH-1211 Geneva, Switzerland}
\affiliation{Group of Applied Physics, University of Geneva, 24 quai Ernest Ansermet, CH-1211
Geneva, Switzerland}
\author{Felix Baumberger}
\email{felix.baumberger@unige.ch}
\affiliation{Department of Quantum Matter Physics, University of Geneva, 24 quai Ernest Ansermet, CH-1211 Geneva, Switzerland}
\affiliation{Swiss Light Source, Paul Scherrer Institute, CH-5232 Villigen, Switzerland}

\begin{abstract}
Black phosphorus (BP) stands out among two-dimensional (2D) semiconductors because of its high mobility and thickness dependent direct band gap. However, the quasiparticle band structure of ultrathin BP has remained inaccessible to experiment thus far. 
Here we use a recently developed laser-based micro-focus angle resolved photoemission ($\mu$-ARPES) system to establish the electronic structure of 2--9 layer BP from experiment. Our measurements unveil ladders of anisotropic, quantized subbands at energies that deviate from the scaling observed in conventional semiconductor quantum wells. We quantify the anisotropy of the effective masses and determine universal tight-binding parameters which provide an accurate description of the electronic structure for all thicknesses. 
\end{abstract}

\maketitle

Few-layer black phosphorus (BP) has an interesting combination of properties and attracts attention for applications in electronics, photonics and sensing~\cite{Li2014b,Youngblood2015,Kim2021}.
In contrast to transition metal dichalcogenides (TMDs), the gap of BP remains direct for any thickness and decreases more rapidly from $\approx 2$~eV in the monolayer down to $\approx 0.3$ ~eV in bulk-like samples spanning the entire technologically relevant mid-infrared range~\cite{Carvalho2016,Tran2014,Qiao2014,Li2017,Yang2015,Zhang2017,Zhang2018}. Light emission from BP is linearly polarized~\cite{Wang2015,Wang2020} and can be tuned by strain~\cite{Rodin2014,Zhang2017a,Kim2021} and gating~\cite{Deng2017,Liu2017,Sherrott2019,Chen2020} over a wide range that includes telecommunications bands. This has been exploited for gas sensing~\cite{Kim2021}, tuneable infrared lasers~\cite{Zhang2020}, variable spectrum detectors and optoelectronic modulators~\cite{Peng2017,Sherrott2019,Kim2021}.

BP is less air-stable than transition metal dichalcogenides (TMDs)~\cite{Island2015,Wang2016} but encapsulation between inert 2D materials was found to be very effective in preventing degradation~\cite{Cao2015,Abate2018}. Few-layer BP devices with carefully protected interfaces achieved low-temperature mobilities far in excess of TMDs~\cite{Chen2015,Long2016,Li2016,Yang2018}. This enabled the first observation of the integer and fractional quantum Hall effect in a 2D material other than graphene~\cite{Yang2018,Li2016}. However, the insight into BP device properties is not yet comparable to graphene, not least because it proved more difficult to establish a tight-binding description of the single particle electronic structure of BP and to determine the relevant parameters.

The electronic structure of BP is more complex than that of few-layer graphene and remains largely unexplored by experiment. 
Monolayer BP consists of puckered honeycomb units, which results in a rectangular unit cell with 4 basis atoms, each contributing 5 electrons distributed in a complex, momentum-dependent way over the 4 $sp$ orbitals of phosphorus~\cite{Morita1986,Rudenko2014,Rudenko2015,Qiao2014}.
It is thus not a-priori clear how to construct a simple, yet accurate effective model of the electronic structure of BP.
Theoretical studies found that the valence band maximum (VBM) of bulk BP has dominant $p_z$ orbital character. This motivated the development of an effective single-orbital tight-binding model of few-layer BP for the relevant states near the Brillouin zone center~\cite{Rudenko2014,Rudenko2015,DeSousa2017}. 
However, because of the puckered structure of BP, even a single orbital model requires an important number of parameters. While a good description of few-layer graphene is obtained with only 2 tight-binding parameters,
theoretical work on BP used 14 parameters to obtain a precise parametrization of first-principles calculations~\cite{Rudenko2015,DeSousa2017}. 
Moreover, such calculations show a significant spread in band gaps, effective masses and subband splittings~\cite{Low2014,Qiao2014,Tran2014,Cai2014,Castellanos-Gomez2014,Rudenko2015,DeSousa2017,Abate2018,Li2017} (see Supporting Information, figures S2, S5) and it is not known how well a certain calculation describes the electronic structure of a real device. This prevented establishing a universal set of tight-binding parameters, which hampered advances in the physics of BP and the exploitation of BP's unique properties in devices.

\begin{figure*}[!htp]
\begin{center}
\includegraphics[width=0.6\textwidth]{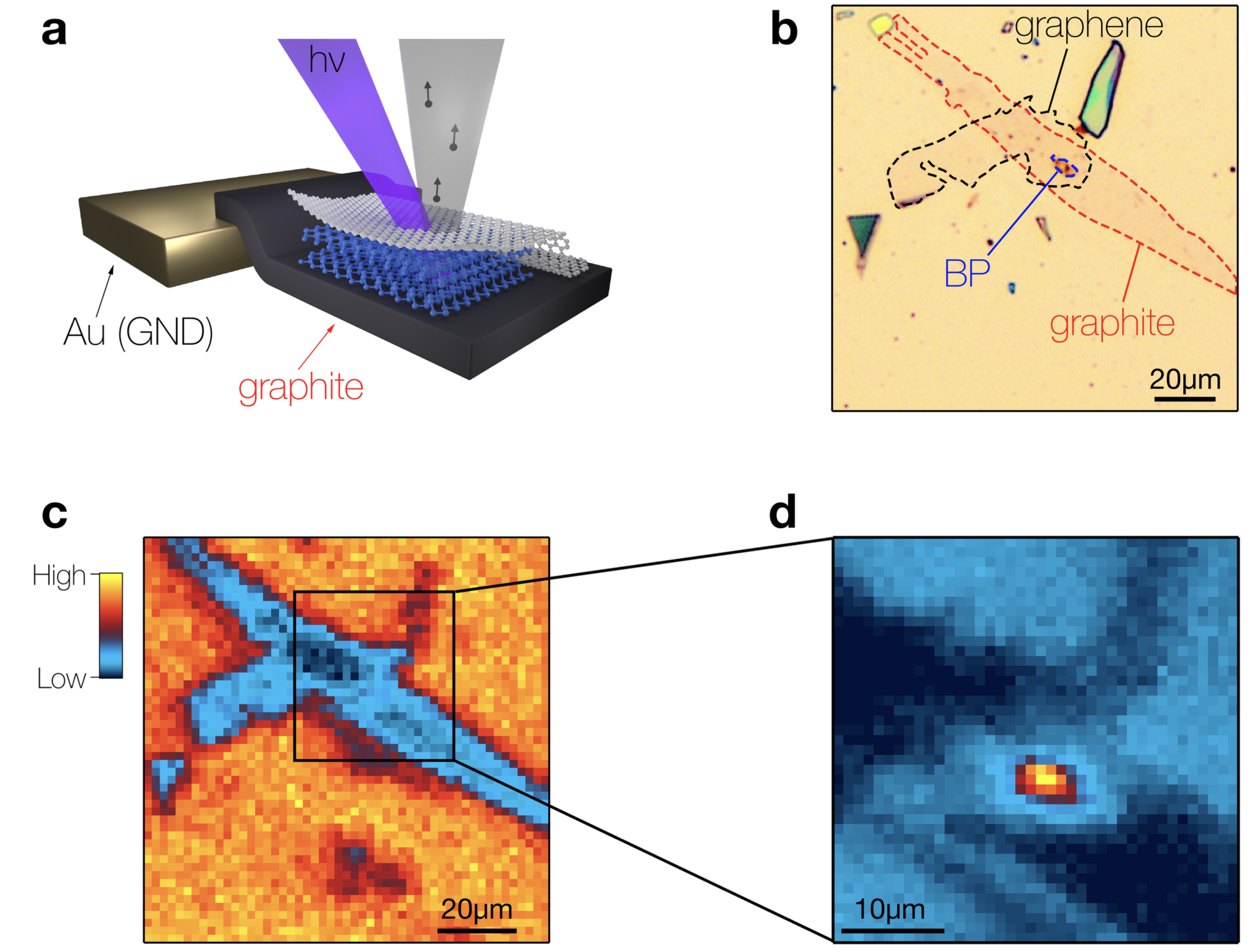}
\caption{Concept of the laser $\mu$-ARPES experiments on few layer BP. (a) Schematic of the micro-focus photoemission setup and of the graphene/BP/graphite heterostructures prepared for this study. (b) Optical micrograph of a $\approx 3 \times 7$~$\mu$m few-layer BP crystal encapsulated in graphite/graphene and supported on a Au-coated Si/SiO$_2$ substrate. (c,d) $\mu$-ARPES photocurrent maps of the device shown in (b) acquired with linear polarizations of the light that minimize (c) and maximize (d) emission from BP.}
\end{center}
\end{figure*}

Here, we use a sensitive laser-based $\mu$-ARPES instrument to accurately map the quantum confined energy bands of $2-9$ layer BP. We quantify the anisotropy and thickness dependence of the effective masses and establish a single set of 8 tight-binding parameters that describes the electronic structure over the full thickness range studied in our experiments.

Figure~1 illustrates the methodology used in our study. We prepare thin crystals of BP by micromechanical exfoliation and use a dry-transfer technique to encapsulate the BP flakes between a graphite bottom electrode and ML graphene. All heterostructures were prepared under protective atmosphere and supported on Au coated Si/SiO$_2$ substrates. More details of the sample preparation are given in Supporting Information, section I. To study the thickness dependence of the electronic structure, we prepared and measured more than a dozen heterostructures.

Typical few-layer BP flakes have lateral dimensions of the order of 10~$\mu$m, far below what is needed for conventional ARPES experiments. To enable electronic structure studies on samples of these dimensions, we developed an instrument that combines the high energy and momentum resolution of laser-ARPES with a spatial resolution below 2~$\mu$m. This is achieved by focusing a 6~eV continuous wave laser (LEOS solutions) with an aberration corrected lens mounted in ultra-high vacuum~\cite{Cucchi2019}. The encapsulated BP flakes are then localized by raster scanning the samples under the focused beam while recording the photocurrent near normal emission in an energy range of a few hundred meV below the Fermi level. In this energy-momentum range, direct transitions are forbidden in graphene and graphite. These materials thus appear as strong depressions in the photocurrent maps, which allows for a convenient correlation with optical micrographs, as illustrated in Figure~1b,c. 
Finally, we identify the BP flake by mapping the photocurrent for an orientation of the linear polarization where BP has a strong photoemission intensity (Figure~1d).

\begin{figure*}[tb]
\begin{center}
\includegraphics[width=0.95\textwidth]{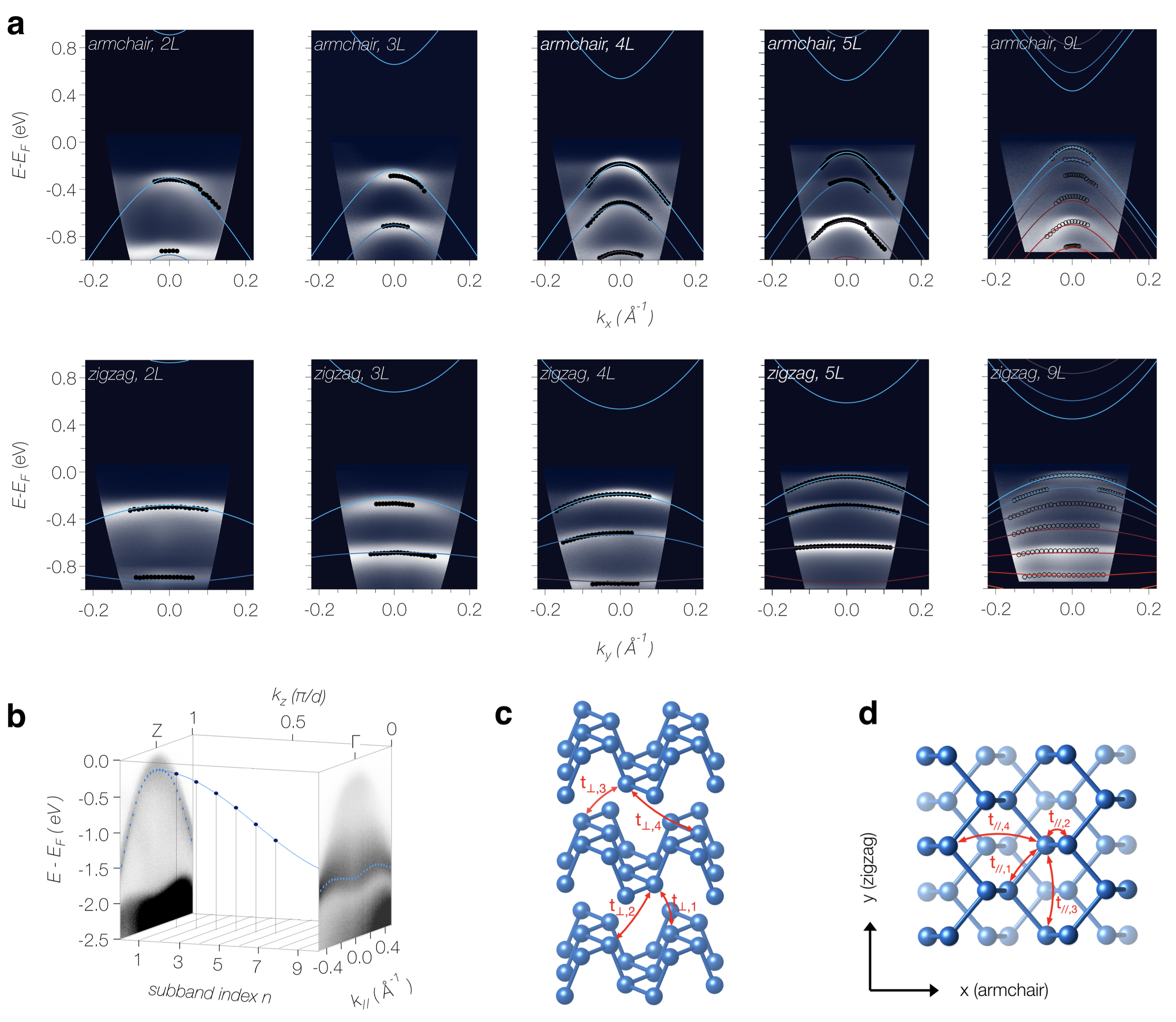}
\caption{Subband dispersion in ultra thin mechanically exfoliated BP. (a) $\mu$-ARPES data of $2-9L$ BP acquired at a sample plate temperature of $T=4.5$~K with linearly polarized 6~eV light (see Methods). Energies are relative to the Fermi level $E_F$.
Solid coloured lines are the result of a global tight-binding fit to subband energies extracted from $2L$, $3L$, $4L$ and $5L$ data (filled black circles). For details, see main text and Supporting Information section III. The same tight-binding model is overlaid on the $9L$ data (open black circles), which were not included in the fit.
(b) Model of the subband energies for $9L$ BP. Gray scale plots show the in-plane dispersion of bulk BP along the zigzag direction measured at $T=50$~K (to avoid charging) and photon energies of $h\nu=20.1$~eV and 36.6~eV to select $k_z$ near the Z- and $\Gamma$ point of the bulk Brillouin zone. Blue symbols indicate the dispersion of the bulk valence band. The measured subband energies for $9L$ BP (black symbols) are well described by the bulk dispersion along $\Gamma$Z and the quantization condition for $k_z$ introduced in the main text. (c,d) Structural model of BP with the 8 tight-binding parameters used in the fit.}
\end{center}
\end{figure*}

Figure~2a shows laser $\mu$-ARPES data from a set of encapsulated BP flakes with $2-9$ layers thickness. The data clearly resolve a series of distinct subbands in the energy range $\sim 1$~eV probed in our experiments. The subband splitting decreases with increasing thickness, as expected for a quantum confined system. However, the subband energies deviate from the scaling with the square of the subband index $n$ observed in conventional semiconductor quantum wells. This is most evident in the $9L$ data, where the subband splitting initially increases with $n$ but then appears to saturate. This behavior can be understood on a semi-quantitative level starting from a phenomenological quantization condition $\frac{n\lambda_n}{2}=(N+\phi)d$ where $N$ is the number of layers, $d$ their thickness and $\phi$ describes the leakage of the wave functions across the interface. 
Since bulk BP has the VBM at the zone boundary $(k_z = \frac{\pi}{d})$, it is convenient to write the quantization of the perpendicular momentum as $k_{z,n}=\frac{\pi}{d}-\frac{2\pi}{\lambda_n}=\frac{\pi}{d}(1-\frac{n}{N+\phi})$ so that the $n=1$ subband defines the VBM.
Using a simple cosine dispersion along $k_z$ -- which is known to be a good approximation to the bulk band structure of BP~\cite{Ehlen2016} -- this model shows excellent agreement with the measured quantum well energies of $9L$ BP for $\phi=1$ and the independently measured bulk bandwidth along $\Gamma$Z $(k_z)$ of 1.35~eV (see Figure~2b). Hence, the above quantization condition provides a simple explanation of a key aspect of the electronic structure. At the same time it serves as a reliable cross-check of the thickness of individual samples (see Supporting Information, Figure~S3).
Describing the conduction band in an analogous way proved to give a good parametrization of the optical gap of BP~\cite{Li2017}. However, such models are not suitable for describing the full quasiparticle dispersion. 

\begin{figure*}[tb]
\centering
\includegraphics[width=0.9\textwidth]{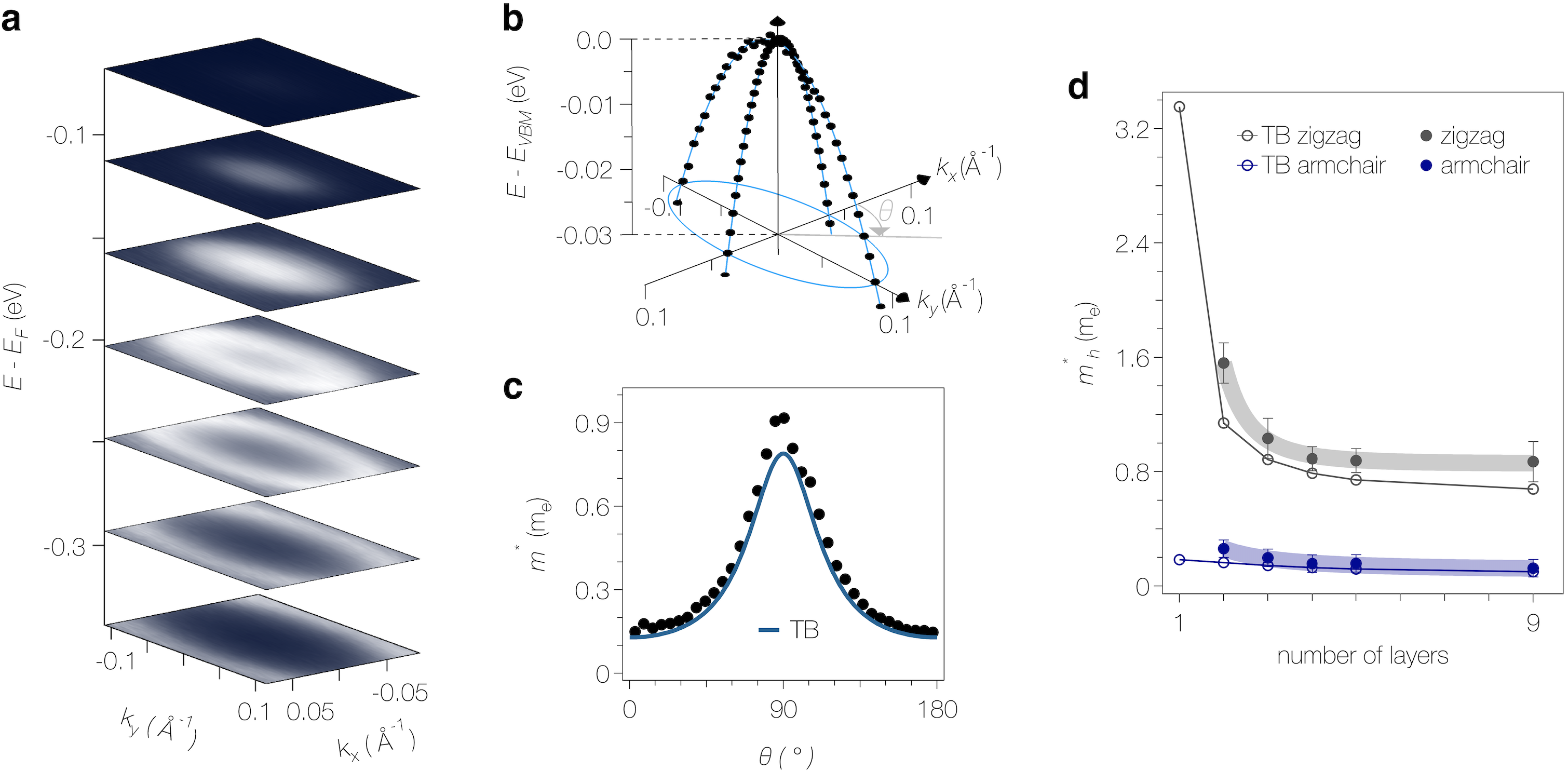}
\caption{Dispersion and effective mass near the valence band edge. (a) Stack of constant energy contours of the top most quantum well state in $4L$ BP. (b) Low-energy band dispersion extracted from curvature plots together with parabolic fits. (c) Anisotropy of the effective mass extracted from the 3D dataset in panel (a) together with the angle dependence of the effective mass obtained from the tight-binding (TB) model. (d) Thickness dependence of the effective masses along both high-symmetry directions. Thick shaded lines are guides to the eye.}
\end{figure*}

BP has a strong structural anisotropy arising from a puckering of the honeycomb lattice of individual layers. DFT studies predicted that this results in a strong and thickness dependent electronic anisotropy. 
Our data in Figure~2 directly confirm a significant electronic anisotropy. Along the zigzag direction, we find a parabolic band with heavy in-plane mass, while the dispersion in the orthogonal armchair direction is stronger and crosses over from parabolic at low energy to nearly linear at high energy.
This dichotomy in the character of holes moving along orthogonal directions can be seen as a remnant of the topological Dirac semimetal state predicted at very high electric field~\cite{Liu2015} and reported in surface-doped bulk BP~\cite{Kim2015b}. 
A careful examination of the data also reveals faint non-dispersive spectral features just below the top of each subband. We will characterize and discuss these features in Figure~4.

First, we quantify effective masses.
Directional effective band masses are of pivotal importance in semiconductor physics but have thus far remained inaccessible to experiment in few-layer BP. Figure~3a shows a series of constant energy contours of $4L$ BP extracted from a full 3D data set. One can clearly recognize elliptical contours over an extended energy range, which is characteristic for an anisotropic two-dimensional electron gas (2DEG) characterized by two effective masses only. The low-energy band dispersion along both high-symmetry directions (Figure~3b) directly confirms a highly parabolic nature of the $n=1$ subband over an energy range of $\approx 30$~meV, corresponding to carrier densities up to $\approx5\cdot10^{12}$~cm$^{-2}$, comparable to the density range accessible with electrostatic gating using hexagonal boron nitride dielectrics. This rationalizes the linear Landau-level splitting reported in Ref~\cite{Yang2018}. 
The effective masses in $4L$ BP are $m^{*}_{x,y}=0.18(3)$ and $0.9(2)$~m$_e$ for the armchair and zigzag direction, respectively. 
This corresponds to a cyclotron mass $m_{c}=\sqrt{m^{*}_{x}m^{*}_{y}}=0.4(1)$~m$_e$, slightly higher than the range of $0.24 : 0.36$~m$_e$ reported in quantum oscillation studies of the charge accumulation layer in gated BP~\cite{Gillgren2015,Li2015,Chen2015,Cao2015,Tayari2015,Yang2018}.
From the above masses of $4L$ BP we obtain an anisotropy $m_y/m_x\approx 4.8$, which is lower than found in most DFT calculations~\cite{Tran2014,Yang2015} but is well reproduced by our tight-binding model (Figure~3c). Extending the analysis of effective masses to other devices reveals a systematic trend to higher effective masses as the thickness is reduced (Figure~3c,d).
This trend is comparable along both high symmetry directions leaving the electronic anisotropy nearly constant.

We now introduce a quantitative tight-binding description of the full experimental quasiparticle dispersion in few-layer BP. 
BP has a complex electronic structure with the 5 valence electrons of phosphorus distributed over all 4 $sp$ orbitals~\cite{Morita1986,Rudenko2014,Rudenko2015,Qiao2014}. However, theoretical studies found that the \textit{ab initio} band structure of few-layer BP near the VBM is well captured by an effective single orbital tight-binding model~\cite{Rudenko2014,Rudenko2015,DeSousa2017}.  
In the following, we determine the parameters of this model directly from experiment. We start our quantitative analysis by extracting band dispersions from a curvature analysis of the experimental data (black circles in Figure~2a)~\cite{Zhang2011}. Studying different parametrizations of the experimental band structure extracted in this way, we find that a minimal model with only 2 in-plane and 1 out-of-plane parameters is insufficient to simultaneously capture the experimentally observed subband splittings and masses. A good description of the data is obtained by restricting the tight-binding model of Refs.~\cite{Rudenko2015,DeSousa2017} to the 4 in-plane and 4 out-of-plane parameters indicated in Figure~2c,d. 
We find that including the additional 6 parameters used in these theoretical works does not improve the fit of our experimental data significantly.
%

\begin{table}[t]
\begin{center}
\begin{tabular}{|>{\centering\arraybackslash}m{1.2cm}|>{\centering\arraybackslash}m{2.2cm}|>{\centering\arraybackslash}m{1.2cm}|>{\centering\arraybackslash}m{2.2cm}|@{}m{0pt}@{}|}
\cline{1-4}
\multicolumn{2}{|c!{\color{white}\vline}}{\cellcolor{black}{\textcolor{white}{intralayer (eV)}}} &\multicolumn{2}{c|}{\cellcolor{black}{\textcolor{white}{interlayer (eV)}}} \\ \hline
\cellcolor[RGB]{220,220,220}$t^{\parallel}_{1}$ & $-1.479$ & \cellcolor[RGB]{220,220,220}$t^{\perp}_{1}$ & $0.552$ \\  \hline
\cellcolor[RGB]{220,220,220}$t^{\parallel}_{2}$ & $3.739$ & \cellcolor[RGB]{220,220,220}$t^{\perp}_{2}$ & $0.115$\\ \hline
\cellcolor[RGB]{220,220,220}$t^{\parallel}_{3}$ & $-0.270$ & \cellcolor[RGB]{220,220,220}$t^{\perp}_{3}$ & $-0.0245$\\ \hline
\cellcolor[RGB]{220,220,220}$t^{\parallel}_{4}$& $0.198$ & \cellcolor[RGB]{220,220,220}$t^{\perp}_{4}$ & $-0.159$\\ \hline
\end{tabular}
\caption{Tight-binding matrix elements determined from a global fit of the measured quasiparticle dispersion of $2L - 5L$ black phosphorus. For further details of the model and a definition of all its parameters see Supporting Information, Section III.}
\label{table:hoppings}
\end{center}
\end{table}

For a robust analysis, we fit the eigenvalues of the 8-parameter tight-binding model
simultaneously to the band dispersions along both high-symmetry directions of all observed quantum well states of the $2L$, $3L$, $4L$ and $5L$ data and additionally restrict the fit to reproduce the bulk band gap. This results in the single set of parameters shown in Table~1. 
In Figure~2a, the bands calculated with this set of parameters are overlaid on the experimental data. We find good agreement with all subband energies as well as the thickness and subband dependent dispersions. Additionally, our tight-binding model also reproduces the thickness dependence of the optical band gap reported in the literature (Supporting Information, Figure~S2)~\cite{Li2017,Yang2015,Zhang2017,Zhang2018}. 
As a further test of the model, we calculate the subband dispersions for $9L$ BP (not included in the fitting procedure) and overlay them on the data in Figure~2a. The good agreement implies that our parametrization provides an accurate description of a wide range of thickness.

\begin{figure}[tb]
\centering
\includegraphics[width=0.48\textwidth]{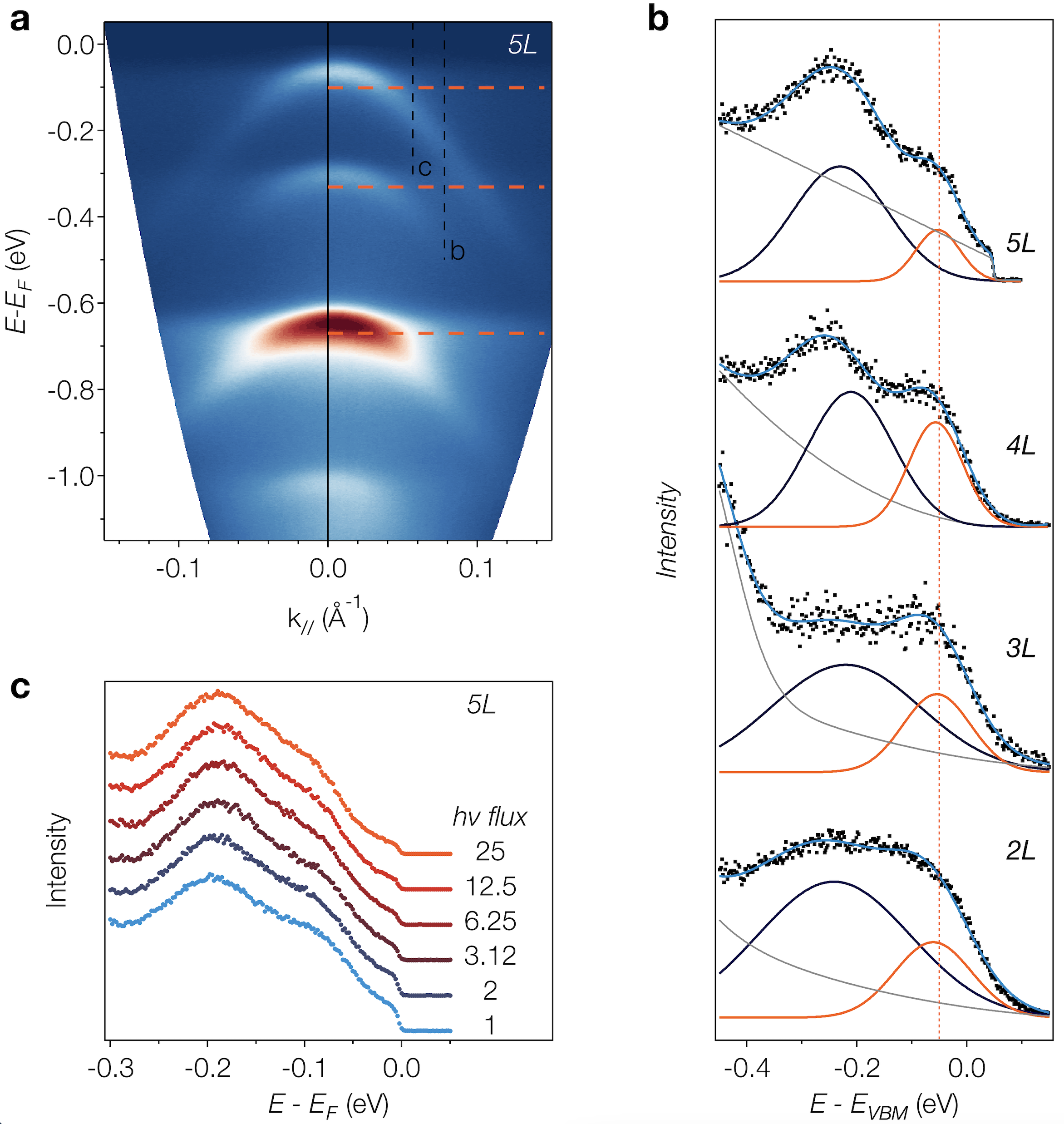}
\caption{Dispersionless spectral features in few-layer BP. (a) ARPES data of $5L$ BP along the armchair direction shown without the normalization of the intensities applied in Figure~2. (b) EDCs of the highest subband in $2-5L$ BP showing a 2-peak structure away from the $\Gamma$ point at $k_{\parallel}\approx 0.08$~\AA$^{-1}$, indicated by a dashed line in (a). (c) EDC at position c indicated in panel (a). The photon flux relative to the first spectrum is indicated in the figure. No significant variation of the spectral line shape is detected when varying the photon flux over more than an order of magnitude.}
\end{figure}

We now discuss the dispersionless spectral signatures discernible along the armchair direction. These features (marked by orange lines in Figure~4a) are ubiquitous in our data. They are present in all devices and are systematically located $\approx 60$~meV below the top of each subband. 
Understanding their origin and contribution to the properties of BP represents a challenge with potential implications for other 2D materials.

An interesting possibility is that these features arise from interface effects. The interface of BP with the graphene encapsulation used in this work forms a periodic moir\'e pattern. Scanning tunnelling microscopy showed that this causes the formation of Landau levels in graphene associated with the strain-induced pseudo magnetic field~\cite{Liu2018}. Such Landau levels are intrinsically non-dispersive. However, their energy scaling and sensitivity to the twist angle~\cite{Liu2018} -- which is uncontrolled in our study and thus inevitably different from device to device -- is inconsistent with our observations.
Flat bands can also arise from a superlattice potential in the active layer and have been detected by ARPES in multiple moir\'e systems~\cite{Utama2021,Lisi2021,Stansbury2021,Gatti2022}. However, the systematic pinning of the dispersionless features at the same energy relative to a subband, irrespective of the twist angle of an individual device, is difficult to reconcile with moir\'e physics. 

We note that a recent time-resolved photoemission study on bulk BP reported Floquet bands with similarities to the dispersionless features in our data~\cite{Zhou2023}. Given the high photon flux density in our measurements ($\approx 10^{14}$~s$^{-1}$ focused in a spot of $\approx 3$~$\mu$m$^2$), one cannot \textit{a priori} exclude light induced effects on the electronic spectrum. This also includes the creation and probing of a steady-state exciton population~\cite{Weinelt2004}.
However, Figure~4c shows that the 2-peak structure arising from the dispersionless features is unaffected if the photon flux is varied by more than an order of magnitude. This excludes Floquet bands~\cite{Zhou2023} or a dynamic exciton population~\cite{Weinelt2004} as the origin of the relevant spectral features.

The 2-peak structure arising from the non-dispersive features also resembles the spectral function of a filled band with strong electron-phonon coupling~\cite{Goodvin2008,Mazzola2013}. Moreover, the energy of $\approx 60$~meV observed experimentally is close to the frequency of optical phonons in BP~\cite{Sohier2018,Neverov2021}.
On the other hand, the high room-temperature mobility of BP indicates that the electron-phonon coupling strength in BP is modest. Simple model calculations for weak to moderate coupling reproduce a dispersionless spectral feature at the right energy but predict that its spectral weight is weak and 
decays more rapidly away from the main band than observed experimentally (see Supporting Information, section~VI).

Interestingly, a recent theoretical study predicts a strong temperature dependence of electron-phonon coupling in BP, raising the possibility of self-trapped small polarons at low-temperature which delocalize at higher temperature~\cite{Neverov2021}.
This might reconcile a high room-temperature mobility with strong electron phonon coupling effects at low temperature.
%
It would be interesting to address this possibility in future temperature-dependent $\mu$-ARPES experiments.
Varying the carrier density by electrostatic gating may provide another route to controlling electron-phonon coupling~\cite{Wang2016b,Nguyen2019}.

In conclusion, we reported a comprehensive study of the quasiparticle dispersion in few-layer BP. Our data directly resolve a multitude of anisotropic quantum well states. We quantify their dispersion and determine universal tight-binding parameters which provide an accurate description of the full electronic structure over a range of thickness. This provides a solid foundation for the interpretation of complementary experiments on BP devices.
We further reported novel dispersionless spectral features which may be relevant for the physical properties of BP.

\section{Acknowledgments}
We thank Alex Ferreira for the development of the dry transfer system.
This work was primarily supported through the Swiss National Science Foundation (SNSF) Div.~II. 
I.G.-L. and A.F.M acknowledge support from the EU graphene flagship.
S.P. acknowledges support from the F.R.S.-FNRS as well as from the EU Horizon 2020 Research and Innovation Programme, under the Marie Sklodowska-Curie Grant Agreement SELPH2D No. 839217 and the PRACE-21 resources MareNostrum at BSC-CNS.
M.G. acknowledges support from the Italian Ministry for University and Research through the Levi-Montalcini program.

\bibliography{thebib.bib}

\end{document}